# Implementación de una Base de Datos Relacional Difusa
# Un Caso en la Industria del Cartón

Leoncio Jiménez♣    Angélica Urrutia♣
José Galindo♦    Pascale Zaraté♥

Resumen

La manipulación de datos, con imprecisión e incertidumbre, es conocida por la comunidad internacional en base de datos, con el término *fuzzy*, que ha sido traducido al español como *difuso* o *borroso*, y al francés como *flou*. En todos estos casos su significado semántico corresponde, por una parte, a la idea natural de *ambiguo* o *vago*, visto del punto vista del razonamiento humano, y por otra parte, a la *teoría de conjuntos difusos*, la *lógica difusa* y la *teoría de la posibilidad* desarrollada por Zadeh entre 1965 y 1977.

El presente artículo da a conocer, por una parte, los atributos del modelo de datos difusos GEFRED (GEneralized model for Fuzzy RElational Databases) [9,10], y por otra parte, la implementación de esos atributos en una Base de Datos Relacional (BDR). El modelado de los atributos fue llevado a cabo en una empresa de fabricación de cartulinas. En este sentido, la fenomenología es descrita a través de atributos y valores imprecisos e inciertos. En particular nos hemos concentrado en el dominio de conocimiento (know-how) relativo al proceso de fabricación de cartulinas estucadas, específicamente al saber hacer del proceso de control de calidad de productos terminados que lleva a cabo el Departamento de Conversión de la empresa. Dado que, para caracterizar la calidad de esos productos, que pueden ser de dos tipos: *pilas* o *rollos*, se utilizan atributos *clásicos* y atributos *fuzzy*. Por lo general, los atributos *clásicos* son medidos con instrumentos físicos, mientras que los atributos *fuzzy*, sólo se pueden apreciar con los sentidos humanos, principalmente la vista y el tacto de los operarios.

**Palabras Clave**: *Modelado de Bases de Datos Relacionales Difusas*.

## 1   Introducción

Los sistemas de información orientados hacia el sector empresarial, nacen a inicios de los años 70, dado por una parte, el crecimiento de la economía, y por otra parte, la necesidad de regular la actividad productiva de la empresa, por medio de la *información*. En este sentido, la teoría general de sistemas, permite describir la *información* a través de un plano mediador entre la *decisión* (conciente de la persona que opera) y la *operación* (sobre datos que describen la realidad). Esta "triada" ha sido identificada, a mediados de los 70, por Jean-Louis Le Moigne, en lo que se conoce desde entonces como modelo OID (Operación, Información, Decisión). Un ejemplo, nos permite ilustrar dicho concepto: "la *decisión* que toma un supermercado en

♣ Dpto. de Computación e Informática, Universidad Católica del Maule, Talca – Chile
 ljimenez@spock.ucm.cl
♣ Dpto. de Computación e Informática, Universidad Católica del Maule, Talca – Chile
 aurrutia@spock.ucm.cl
♦ Dpto. de Lenguajes y Ciencias de la Computación, Universidad de Málaga, Málaga – España
ppgg@lcc.uma.es
♥ IRIT, UMR 5505 CNRS – INPT – ENSIACET – GI, Toulouse – France  Pascale.Zarate@irit.fr

aumentar su nivel de stock (*operación*), digamos de 300 a 1000 unidades de un determinado producto (dato), es regulada por la *información externa*, que el sistema de información, percibe y distingue de su entorno (por ejemplo, los efectos del lanzamiento al mercado de una nueva campaña publicitaria), y por otra parte, la *información interna*, que el sistema de información mantiene en sus *bases de datos* (por ejemplo, los niveles de stock en un momento determinado del inventario).

Una Base de Datos (BD), en un contexto genérico, debe permitir dos operaciones básicas. Una de ellas es almacenar datos, y la otra es consultar datos, para ello existen variadas herramientas de diseño y lenguajes para modelar, diseñar e implementar una BD, según sea su tipo (relacional, objeto, distribuida, etc.).

Este artículo, se centra en las Bases de Datos Relacionales (BDR), es decir, una base de datos bajo un modelo relacional.

Los tipos de datos contenidos en la BDR, generalmente, se denominan *clásicos*, porque en su representación, son datos precisos (por ejemplo: 41), datos desconocidos (valor desconocido pero el atributo es aplicable), no definidos (atributo no aplicable o sin sentido) o nulos (ignorancia total, no sabemos nada sobre eso). Mientras, en lo que se refiere a la consulta son datos sin incertidumbre, es decir el resultado de un SELECT a una tabla de la BDR, se trata de una tupla precisa de valores. Sin embargo, estos tipos de datos (precisos, desconocidos, no definidos y nulos) no nos permiten describir fenómenos que manifiesten cierta imprecisión y/o incertidumbre, tanto en su representación como en su consulta.

En Francia, a principios de los 80, surge uno de los primeros estudios matemáticos sobre el tratamiento de la información difusa, es decir, la información que encierra alguna imprecisión o incertidumbre en una BDR. Estos fueron realizados, simultáneamente, por Dubois y Prade, en dos tesis doctorales [2,10], a partir de los trabajos sobre la incertidumbre de Zadeh, específicamente, la teoría de la posibilidad, que tiene sus raíces en otras dos investigaciones de Zadeh: la teoría de conjuntos difusos[1] y la lógica difusa. Sin embargo, no es hasta 1984, que la tesis doctoral de Testemale (dirigida por Prade) [12], propone un modelo de datos difusos, para la implementación de una BDRD en una BD relacional. Dicho modelo se conoce, desde entonces, como el modelo de Prade-Testemale [3].

Paralelamente, surgen otros modelos de datos difusos para la implementación de una BDRD, en particular: el modelo de Umano-Fukami, el modelo de Buckles-Petry, el modelo de Zemankova-Kaendel, y el modelo GEFRED de Medina [5,9,13].

Es justamente, la tipología de los atributos difusos del modelo GEFRED que serán explicados en este artículo, como también su implementación en una BDR, por medio de la FMB (Fuzzy Metaknowledge Base).

## 2   Modelos de datos difusos

Algunos investigadores utilizan los modelos de datos difusos, por una parte, para representar atributos difusos mediante la teoría de conjuntos difusos, y por otra parte, para interrogar una BDRD mediante una BDR.

---

[1] Por algunos también llamada *teoría de subconjuntos difusos* [8] por el hecho que el universo del discurso es un conjunto definido por la *teoría de conjuntos*.

A nivel conceptual, en [6,13] se propone una extensión del modelo EER (Enhanced Entity Relationship), para representar los atributos difusos del modelo GEFRED, así como también una herramienta CASE, llamada FuzzyCASE [13,14] que los soportan.

A nivel lógico, en [1,4] se proponen cuantificadores, comparadores y grados difusos para manipular los atributos difusos del modelo GEFRED en la consulta, bajo una plataforma Oracle 8.

## 2.1   Teoría de conjuntos difusos

La teoría de conjuntos difusos [15] ha sido propuesta por Zadeh desde 1965, a partir del concepto de conjunto (una colección de objetos). El supuesto de esta teoría es que existen conjuntos en los que no está claramente determinado si un elemento pertenece o no al conjunto. En general, un elemento pertenece al conjunto con cierto grado. Lo anterior, en lenguaje matemático, es como sigue:

Un conjunto difuso $A$, sobre un universo de discurso $U$ es un conjunto de pares, dado por:

$$A = \{\mu_A(u)/u: u \in U, \mu_A(u) \in [0,1]\},$$

Donde, $\mu$ es llamada *función de pertenencia* y $\mu_A(u)$ es el *grado de pertenencia* del elemento $u$ al conjunto difuso $A^2$. Este grado de pertenencia oscila entre los extremos 0 y 1, donde:

$\mu_A(u) = 0$, indica que $u$ no pertenece en absoluto al conjunto difuso $A$,
$\mu_A(u) = 1$, indica que $u$ pertenece totalmente al conjunto difuso $A$.

**Ejemplo 1**

La edad de una persona es un atributo que la caracteriza, entonces el concepto "joven", puede ser representado por un conjunto difuso, de la forma: Joven = {0/15, 1/20, 1/25, 0/30}. Esto, quiere decir, que las personas de 15 ó 30 años no son jóvenes, pero si lo son las personas de 20 ó 25 años. En tanto, que una persona de 26 años lo es con grado 0.8. En este caso el conjunto difuso es representado por *valores numéricos*.

**Ejemplo 2**

El conjunto de las personas que son "altas" es un conjunto difuso, pues no está claro el límite de altura que se establece a partir de que medida una persona es alta o no lo es. Ese límite es difuso y, por lo tanto, el conjunto que lo delimita también lo será. En este caso, el conjunto difuso es representado por *escalares simples*.

---

[2] Esa misma función $\mu$ en la teoría de posibilidad es llamada *distribución de posibilidad*, mientras $\mu_A(u)$ es llamado *grado de posibilidad*. Este artículo utiliza las dos, dependiendo del contexto.

## 2.2 Datos difusos de una BDRD

La teoría de conjuntos difusos [15], en particular la teoría de posibilidad permite definir dos tipos de datos difusos de una BDRD, estos son:

**Etiqueta lingüística**

Una *etiqueta lingüística* es aquella palabra, en lenguaje natural, que representa un conjunto difuso. Este conjunto puede estar formalmente definido (la etiqueta "joven" en el ejemplo 1) o no (la etiqueta "alto" en el ejemplo 2). En este sentido, otras etiquetas lingüísticas que permiten describir fenómenos que manifiestan cierta imprecisión, son: "viejo", "frío", "caliente", "templado", "barato", "caro", "bajo", "grande", "pequeño", etc.

Es bueno hacer notar. Primero, que la definición intuitiva de esas etiquetas, no sólo puede variar de una persona a otra y del momento particular, sino que también puede variar, según el contexto en que se aplique. En efecto, no nos referimos a la misma altura cuando se trata de una persona "alta", que de un edificio "alto". Segundo, estas etiquetas lingüísticas son definidas sobre un dominio con referencial ordenado. Esto permite representar la imprecisión de los atributos con etiquetas lingüísticas. Veamos esto con un ejemplo.

**Ejemplo 3**

Supongamos que sobre el atributo "edad", además de la etiqueta lingüística "joven", se encuentra definidas otras dos etiquetas: "maduro" y "mayor". La figura 1, muestra el dominio de la *función de pertenencia* para cada una de esas etiquetas. Así, la edad 26 tiene un grado de pertenencia 0.8 para la etiqueta lingüística "joven". Gráficamente, la función de pertenencia, corresponde a un trapecio dado por sus 4 valores característicos, por ejemplo: {0/25, 1/30, 1/40, 0/45}, para la etiqueta lingüística "maduro".

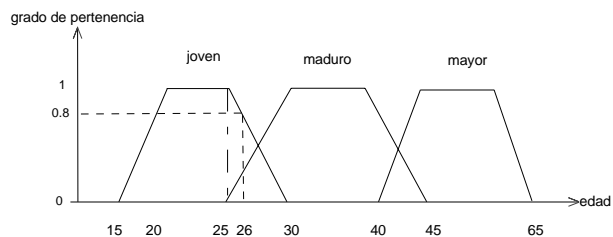

**Fig. 1** Función de pertenencia de las etiquetas lingüísticas del atributo "edad"

Cómo muestra la figura 1, el atributo "edad" está definido sobre un dominio con referencial ordenado, lo que permite comparar los valores de las etiquetas. En caso, que el referencial sea no ordenado, es necesario el uso de una función de similitud.

**Relación de similitud**

Una *relación de similitud* permite comparar etiquetas lingüísticas definidas en dominios con referencial no ordenado. Esta relación indica, que para cada dominio, es necesario definir una *función de similitud* que permita medir la similitud o parecido entre dos valores del dominio. Matemáticamente, una función de similitud $s_r$ puede ser vista como el producto cartesiano sobre dos dominios $D$, tal que:

$$s_r: D \times D \rightarrow [0,1]$$
$$s_r(d_i, d_j) \rightarrow [0,1] \text{ con } d_i, d_j \in U \text{ con } i, j \in R$$

Los valores $d_i$, $d_j$ (de similitud) están normalizados en un intervalo [0,1], correspondiendo el 0 al significado "totalmente diferente" y el 1 al significado "totalmente parecido" (o iguales). El dominio *D*, puede estar definido para un conjunto de asignaciones excluyentes de valores numéricos o escalares simples, por una parte, con $\mu_A(u) = 1$, y por otra parte, con $\mu_A(u) \in [0,1]$.

**Ejemplo 4**

Supongamos que tenemos el atributo "color de pelo", donde el dominio se encuentra definido por tres etiquetas lingüísticas: "Rubio", "Moreno", y "Pelirrojo". Es este caso el conjunto difuso definido por esas etiquetas, no se encuentra definido en un referencial ordenado, lo que hace necesario la implementación de una *función de similitud*, que por comodidad, normalmente se presenta en un formato de matriz, tal como lo muestra tabla 1.

| $s_r$ | Rubio | Moreno | Pelirrojo |
|---|---|---|---|
| Rubio | 1 | 0.1 | 0.8 |
| Moreno | 0.1 | 1 | 0.3 |
| Pelirrojo | 0.8 | 0.3 | 1 |

**Tabla 1** Función de similitud para atributo "color de pelo"

La tabla 1, muestra una relación de semejanza entre las etiquetas. En otras palabras, los valores indicados en ella, muestra en "qué se parece" una etiqueta con otra. Por ejemplo, entre un color de pelo rubio y moreno la función de similitud vale 0.1. En general, el conjunto difuso definido para la etiqueta lingüística "Rubio" es: {1/Rubio, 0.1/Moreno, 0.8/Pelirrojo}.

## 3   El modelo GEFRED

El modelo relacional difuso GEFRED (GEneralized model for Fuzzy RElational Databases) [9,10] permite representar los siguientes atributos difusos:

- ➢ **Atributo difuso de Tipo 1:** Estos atributos son utilizados para representar "*valores precisos*". El dominio *D*, asociado a este tipo de atributo puede estar definido por (a) valores numéricos; y (b) por escalares simples, en ambos casos, el grado de posibilidad es 1, es decir con $\mu_A(u) = 1$.

- ➢ **Atributo difuso de Tipo 2:** Estos atributos son utilizados para representar "*valores imprecisos sobre referencial ordenado*". El dominio *D*, asociado a este tipo de atributo puede estar definido por (a) valores numéricos; (b) por escalares simples, en ambos casos, con grado de posibilidad 1; (c) por etiquetas lingüísticas; (d) por valores parcialmente desconocidos entre dos valores precisos; (e) por valores parcialmente desconocidos entre un valor preciso, en esos tres casos, con grado de posibilidad entre 0 y 1, es decir, con $\mu_A(u) \in [0,1]$; (f) por valores desconocidos con grado de posibilidad 1; (g) por valores inaplicables con grado de posibilidad 0; y (h) por valores nulos con grado de posibilidad 1.

- ➢ **Atributo difuso de Tipo 3:** Estos atributos son utilizados para representar "*valores imprecisos sobre referencial no ordenado*". El dominio *D*, asociado a este tipo de atributo puede estar definido por (a) valores excluyentes de números o escalares con grado de posibilidad 1, es decir, con $\mu_A(u) = 1$; (b) por relaciones de similitud con grado de posibilidad entre 0 y 1, es decir, con $\mu_A(u) \in [0,1]$; (c) por valores desconocidos con grado de posibilidad 1; (d) por valores inaplicables con grado de posibilidad 0; y (e) por valores nulos con grado de posibilidad 1.

Estos atributos son representados, como veremos a continuación, por los siguientes conjuntos difusos:

**CRISP** (valores numéricos de atributos difusos Tipo 1 o Tipo 2, ver ejemplo 1), **LABEL** (escalares simples de atributos difusos Tipo 1 o Tipo 2, ver ejemplo 2), **TRAPECIO** (etiquetas lingüísticas de atributos difusos Tipo 2, ver ejemplo 3), **INTERVALO** (valores parcialmente desconocidos entre dos valores precisos de atributos difusos Tipo 2), **APROXIMADAMENTE** (valores parcialmente desconocidos entre un valor preciso de atributos difusos Tipo 2), **UNKNOWN** (valores desconocidos de atributos difusos Tipo 2 o Tipo 3), **UNDEFINED** (valores inaplicables de atributos difusos Tipo 2 o Tipo 3), **NULL** (valores nulos de atributos difusos Tipo 2 o Tipo 3), **SIMPLE** (valores excluyentes de números o escalares de atributos difusos Tipo 3), **DISTRIBUCION POSIBILIDAD** (relaciones de similitud de atributos difusos Tipo 3, ver ejemplo 4).

## 4  Implementación de los atributos Tipo 1, 2 y 3 en la FMB

La FMB (Fuzzy Metaknowledge Base) [9], es una extensión del catálogo del SGBD relacional, que permite (1) almacenar los datos difusos en la BDR, y (2) transformar las consultas FSQL en SQL.

FSQL (Fuzzy SQL) [9], tanto a nivel cliente como servidor, es el protocolo de comunicación con el usuario o el SGBD relacional, que permite extender SQL para el tratamiento de datos difusos. Así, toda sentencia FSQL es traducida por el Servidor FSQL al lenguaje SQL de modo que su sintaxis es reconocida por el SGBD relacional. Mientras, que el SGBD, realiza la consulta a la base de datos (entendiéndose por consulta a operaciones como: select, insert, update, delete, etc.), ya sea, empleando el lenguaje SQL o FSQL.

### 4.1  Atributos difusos Tipo 1, 2 y 3 en la FMB

La FMB contiene los siguientes atributos difusos:

**a) CRISP**

El conjunto difuso CRISP es definido por una distribución de posibilidad sobre un intervalo real [0,1] de la forma:

$$\text{CRISP} = \{\mu_{\text{CRISP}}(x)/x : x \in U\}$$

En que, para todo número simple $x$ de un atributo difuso Tipo 1 o Tipo 2, en un dominio ordenado $U$, se tiene: $\mu_{\text{CRISP}}(x) = 1$.

Luego,               CRISP = {1/CRISP}

CRISP es utilizado para representar atributos difusos Tipo 1 o Tipo 2, sobre referencial ordenado, cuyo valor está dado por un *valor numérico*.

Ejemplo: Sea "Edad" un atributo difuso Tipo 1 o Tipo 2. Supongamos Edad = 28 (un número simple cualquiera). Entonces, CRISP es representado mediante una distribución de posibilidad, de la forma: CRISP = {1/28}.

**b) LABEL**

El conjunto difuso LABEL es definido por una distribución de posibilidad sobre un intervalo real [0,1] de la forma:

$$LABEL = \{\mu_{LABEL}(x)/x : x \in U\}$$

En que, para todo escalar simple $x$ de un atributo difuso Tipo 1 o Tipo 2, en un dominio ordenado $U$, se tiene: $\mu_{LABEL}(x) = 1$.

Luego,               LABEL = {1/LABEL}

LABEL es utilizado para representar un atributo difuso Tipo 1 o Tipo 2, sobre referencial ordenado, cuyo valor está dado por un *escalar simple*.

Ejemplo: Sea "Tamaño" un atributo difuso Tipo 2. Supongamos Edad = Grande (un escalar simple cualquiera). Entonces, LABEL es representado mediante una distribución de posibilidad, de la forma: LABEL = {1/Grande}.

**c) TRAPECIO**

El conjunto difuso TRAPECIO es definido por una distribución de posibilidad sobre un intervalo real [0,1] de la forma:

$$TRAPECIO = \{\mu_{TRAPECIO}(x)/x : x \in U=[a,b,c,d], \mu_{TRAPECIO}(x) \in [0,1]\}$$

En que, para todo valor trapecio $x$ de un atributo difuso Tipo 2, en un dominio ordenado $U$, caracterizado por el intervalo $[a,b,c,d]$ con $a<b<c<d$ se tiene:

$$\mu_{TRAPECIO}(x) = \begin{cases} [0,1] & \text{si } a \leq x \leq b \\ 1 & \text{si } b \leq x \leq c \\ [0,1] & \text{si } c \leq x \leq d \end{cases}$$

Luego,               TRAPECIO = {[0,1]/ $a \leq x \leq b$, 1/$b \leq x \leq c$, [0,1]/ $c \leq x \leq d$}

TRAPECIO es utilizado para representar un atributo difuso Tipo 2, sobre referencial ordenado, cuyo valor está dado por una *etiqueta lingüística*.

Ejemplo: Ver figura 1.

**d) UNKNOWN**

El conjunto difuso UNKNOWN es definido por una distribución de posibilidad sobre un intervalo real [0,1] de la forma:

$$UNKNOWN = \{\mu_{UNKNOWN}(x)/x : x \in U\}$$

En que, para todo valor desconocido $x$ de un atributo difuso Tipo 2 o Tipo 3, en un dominio ordenado $U$, se tiene: $\mu_{UNKNOWN}(x) = 1$.

Luego, $\quad UNKNOWN = \{1/UNKNOWN\}$

UNKNOWN es utilizado para representar atributos difusos Tipo 2 o Tipo 3, sobre referencial ordenado y no ordenado respectivamente, donde el valor existe y el dominio del atributo es aplicable al objeto, pero el valor es completamente desconocido. Es decir, hay una ignorancia total en relación al valor que toma un atributo difuso Tipo 2 o Tipo 3.

En este caso, el sistema de información tiene un conocimiento imperfecto del valor del atributo para un objeto dado, es decir, los valores del atributo son mal conocidos.

**e) UNDEFINED**

El conjunto difuso UNDEFINED es definido por una distribución de posibilidad sobre un intervalo real [0,1] de la forma:
$$UNDEFINED = \{\mu_{UNDEFINED}(x)/x : x \in U\}$$

En que, para todo valor inaplicable $x$ de un atributo difuso Tipo 2 o Tipo 3, en un dominio ordenado $U$, se tiene: $\mu_{UNDEFINED}(x) = 0$.

Luego, $\quad UNDEFINED = \{0/UNDEFINED\}$

UNDEFINED es utilizado cuando ningún valor es posible para un atributo difuso Tipo 2 o Tipo 3. Es decir, El valor $x$ existe y es conocido, pero el dominio $U$ del atributo es inaplicable al objeto.

En este caso, el sistema de información tiene un conocimiento imperfecto del valor del atributo para un objeto dado, es decir, los valores de los atributos difusos Tipo 2 o Tipo 3 no puede tomar ningún valor en el dominio ordenado $U$.

**f) NULL**

El conjunto difuso NULL es definido por una distribución de posibilidad sobre un intervalo real [0,1] de la forma:

$$NULL = \{\mu_{NULL}(x)/x : x \in UNKNOWN, x \in UNDEFINED\}$$

En que, para cada valor nulo $x$ de un atributo difuso Tipo 2 o Tipo 3, en un dominio ordenado $U$, se tiene: $\mu_{NULL}(x) = 1$.

Luego,            NULL = {1/UNKNOWN, 1/UNDEFINED}

**g) INTERVALO**

El conjunto difuso INTERVALO es definido por una distribución de posibilidad sobre un intervalo real [0,1] de la forma:

$$\text{INTERVALO} = \{\mu_{\text{INTERVALO}}(x)/x : x \in U=[n,m], \mu_{\text{INTERVALO}}(x) \in [0,1]\}$$

En que, para todo valor $x$ de un atributo difuso Tipo 2, en un dominio ordenado $U$, caracterizado por el intervalo = $[n,m]$ con $n<m$, es decir, dos valores precisos cualquiera, se tiene:

$$\mu_{\text{INTERVALO}}(x) = \begin{cases} 0 & \text{si } m < x \\ 1 & \text{si } n \leq x \leq m \\ 0 & \text{si } x < n \end{cases}$$

Luego,      INTERVALO = {0/ $x < n$, 1/ $n \leq x \leq m$, 0/ $m < x$}

**h) APROXIMADAMENTE**

El conjunto difuso APROXIMADAMENTE es definido por una distribución de posibilidad sobre un intervalo real [0,1] de la forma:

$$\text{APROXIMADAMENTE} = \{\mu_{\text{APROXIMADAMENTE}}(x)/x : x \in U=[d-\gamma, d+\gamma], \mu_{\text{APROXIMADAMENTE}}(x) \in [0,1]\}$$

En que, para todo valor aproximado $x$ de un atributo difuso Tipo 2, en un dominio ordenado $U$, caracterizado por el intervalo = $[d-\gamma, d+\gamma]$, donde $d$ es el valor aproximado y $\gamma$ es un margen para todo elemento $x$ del dominio $U$, se tiene:

$$\mu_{\text{APROXIMADAMENTE}} = \begin{cases} [0,1] & \text{si } d-\gamma \leq x \\ 1 & \text{si } x = d \\ [0,1] & \text{si } x \leq d+\gamma \end{cases}$$

Luego,     $\mu_{\text{APROXIMADAMENTE}} = \{[0,1]/\ d-\gamma \leq x,\ 1/d=x,\ [0,1]/\ d+\gamma \geq x\}$

**i) SIMPLE**

El conjunto difuso SIMPLE es definido por una distribución de posibilidad sobre un intervalo real [0,1] de la forma:
$$\text{SIMPLE} = \{\mu_{\text{SIMPLE}}(x)/x : x \in U\}$$

En que, para todo conjunto de asignaciones excluyentes de números o escalares *x* de un atributo difuso Tipo 3, en un dominio no ordenado *U*, se tiene: $\mu_{SIMPLE}(x) = 1$.

Luego, SIMPLE = {1/SIMPLE}

SIMPLE es utilizado para representar atributos difusos Tipo 3, sobre referencial no ordenado, donde el valor es un número o un escalar.

Ejemplo: Sea "Edad" un atributo difuso Tipo 3. Supongamos Edad = {27, 28}. Entonces, SIMPLE es representado mediante una distribución de posibilidad, de la forma: SIMPLE = {1/27, 1/28}. En caso de tener escalares, por ejemplo, Edad = {JOVEN, ADULTO}. Entonces, SIMPLE es representado mediante una distribución de posibilidad, de la forma: SIMPLE = {1/JOVEN, 1/ADULTO}.

### j) DISTRIBUCION POSIBILIDAD

El conjunto difuso DISTRIBUCION POSIBILIDAD es definido por una distribución de posibilidad sobre un intervalo real [0,1] de la forma:

$$\text{DISTRIBUCION POSIBILIDAD} = \{\mu_{DISTRIBUCION\ POSIBILIDAD}(x)/x : x \in U\}$$

En que, para todo conjunto de posibles asignaciones excluyentes de números o escalares *x* de un atributo difuso Tipo 3, en un dominio no ordenado *U*, se tiene: $\mu_{DISTRIBUCION\ POSIBILIDAD}(x) \in [0,1]$.

Luego, $\mu_{DISTRIBUCION\ POSIBILIDAD}(x) = \{[0,1]/x\}$

DISTRIBUCION POSIBILIDAD es utilizado para representar atributos difusos Tipo 3, sobre referencial no ordenado, en el dominio de los números o de los escalares.

Ejemplo: Sea "Edad" un atributo difuso Tipo 3. Supongamos Edad = {27, 28, 29}. Entonces, DISTRIBUCION POSIBILIDAD es representado mediante una relación de similitud, de la forma: SIMPLE = {0.4/27, 1/28, 0.8/29}. En caso de tener escalares, por ejemplo, Edad = {JOVEN, ADULTO}. Entonces, DISTRIBUCION POSIBILIDAD es representado mediante una relación de similitud, de la forma: DISTRIBUCION POSIBILIDAD = {0.6/JOVEN, 1/ADULTO}.

## 4.2  Tablas de conversión de los atributos difusos Tipo 1, 2 y 3 en la FMB

Los conjuntos difusos Tipo 1, 2 y 3, de la forma: CRISP, LABEL, TRAPECIO, INTERVALO, APROXIMADAMENTE, UNKNOWN, UNDEFINED, NULL, SIMPLE, DISTRIBUCION POSIBILIDAD, son almacenadas en la FMB a través de *tablas de conversión*.

### 4.2.1  Tabla de conversión para atributos difusos Tipo 1

Para este tipo de atributo no existe una tabla de conversión propiamente tal en la FMB, ya que los valores numéricos (CRISP) y escalaras simples (LABEL) reciben una representación igual que los datos precisos.

### 4.2.2 Tabla de conversión para atributos difusos Tipo 2

La tabla 2, muestra el protocolo de conversión que la FMB utiliza para atributos Tipo 2. La primera columna muestra los atributos Tipo 2, que son posibles de almacenar en la FMB. La segunda columna muestra el identificador asociado a cada uno de ellos. Por ejemplo, el conjunto difuso APROXIMADAMENTE tiene asociado el identificador 6. La tercera columna, se subdivide en cuatro columnas, cada una de ellas almacena los valores de las variables que caracterizan ese conjunto. La primera columna almacena el valor del dato (d), la segunda columna su límite izquierdo (d-margen), la tercera columna su límite derecho (d+margen), y la cuarta columna el margen.

|  | Id | V1 | V2 | V3 | V4 |
|---|---|---|---|---|---|
| UNKNOWN | 0 | *null* | *null* | *null* | *null* |
| UNDEFINED | 1 | *null* | *null* | *null* | *null* |
| NULL | 2 | *null* | *null* | *null* | *null* |
| CRISP | 3 | d | *null* | *null* | *null* |
| LABEL | 4 | FUZZY_ID | *null* | *null* | *null* |
| INTERVALO | 5 | n | *null* | *null* | m |
| APROXIMADAMENTE | 6 | d | d-margen | d+margen | margen |
| TRAPECIO | 7 | α | β-α | γ-δ | δ |

**Tabla 2** Protocolos de conversión en la FMB para atributos Tipo 2[3]

### 4.2.3 Tabla de conversión para atributos difusos Tipo 3

La tabla 3, muestra el protocolo de conversión que la FMB utiliza para atributos difusos Tipo 3. La primera columna muestra las etiquetas que es posible almacenar para un atributo Tipo 3. La segunda columna muestra el identificador asociado a cada etiqueta. Por ejemplo, el conjunto difuso DISTRIBUCION POSIBILIDAD, tiene asociado el identificador 4. La tercera columna, se subdivide en *n* columnas de *n* parejas, con $n \geq 1$, (FP1, F1),..., (FPn, Fn), donde es posible almacenar los valores de la distribución de posibilidad. Estos valores se encuentran en el intervalo [0, 1]. Por ejemplo, la DISTRIBUCION POSIBILIDAD de la tabla 1, estaría formada por 9 parejas: (1/Rubio), (0.1/Moreno), (0.8/Pelirrojo), (0.1/Rubio), (1/Moreno), (0.3/Pelirrojo), (0.8/Rubio), (0.3/Moreno), (1/Pelirrojo).

|  | FT | FP1 | F1 | ... | $FP_n$ | $F_n$ |
|---|---|---|---|---|---|---|
| UNKNOWN | 0 | *null* | *null* | ... | *null* | *null* |
| INDEFINED | 1 | *null* | *null* | ... | *null* | *null* |
| NULL | 2 | *null* | *null* | ... | *null* | *null* |
| SIMPLE | 3 | p | d | ... | *null* | *null* |
| DISTRIBUCION POSIBILIDAD | 4 | $p_1$ | $d_1$ | ... | $p_n$ | $d_n$ |

**Tabla 3** Protocolos de conversión en la FMB para atributos Tipo 3

---

[3] Los valores *null* de las tablas 2 y 3, corresponden al null de una BDR, para indicar valores no aplicables.

## 5   Caso de estudio

El caso de estudio propuesto en este artículo, tiene relación con el control de calidad de una empresa de fabricación de cartulinas, uno de los problemas en la Bodega del Departamento de Conversión (ver figura 2), es el estado de los productos terminados, por ejemplos, "sucios" o "húmedos". En dicho escenario, por ejemplo, un requerimiento válido en lenguaje natural, que podría implementarse en FSQL, sería: "*Obtenga los rollos de cartulina almacenados en la bodega que estén sucios o húmedos*"[4]. Es claro, que estos criterios de búsqueda pueden ser considerados como ambiguos o vagos, ya que lo que es sucio para un operario, no lo es para otro. Por otra parte, en nuestro modelo hemos considerado los parámetros: Formato_Largo, Formato_Ancho (diámetro) y Altura de los *rollos* como difusos, solamente para ilustrar nuestra propuesta, pero en la realidad (formulario de pedidos) no lo son. Lo mismo, los atributos Formato_Largo (altura), Formato_Ancho de las *pilas*. Por ejemplo, si consideramos el atributo Formato_Ancho (diámetro) de un *rollo*, en una BDR sólo se aceptarán valores exactos del diámetro, por ejemplo: 400 cm. Ahora, que sucede si 400 cm. es la norma especificada para el *rollo*, y se tienen almacenados otros *rollos* en la Bodega del Departamento de Conversión (ver figura 2), cuyos diámetros están por encima o por debajo de lo especificado por la norma, el problema parece evidente: ¿Cómo clasificar esos *rollos* en el sistema?. Es justamente en ese tipo de situación que las BDRD se vuelven interesantes, dado que es posible definir unas *etiquetas lingüísticas* dada por el conjunto difuso: {"debajo-de-la-norma", "en-la-norma", "arriba-de-la-norma"}[5], en que la pertenencia de un valor, por ejemplo: 402 cm, a una de esas *etiquetas lingüísticas*, por ejemplo, "debajo-de-la-norma" y "en-la-norma", pueda ser incierta. Es decir, el valor 402 pertenece a esas dos *etiquetas lingüísticas* (o subconjuntos difusos), pero con un *grado de pertenencia*, por ejemplo 0,4 y 0,6, respectivamente.

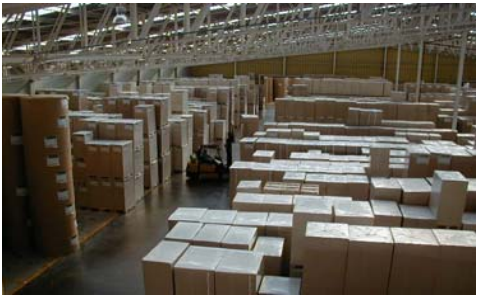
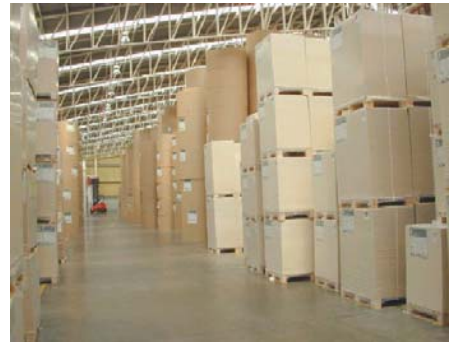

**Fig. 2** Bodega de productos terminados (Pilas y Rollos) del Departamento de Conversión

Los atributos difusos de los productos terminados (*pilas* o *rollos*), se refieren a las características que deben tener dichos productos. Ahora bien, estas características no son medidas con instrumentos físicos, ya que sólo se pueden apreciar con los sentidos humanos, principalmente la vista y el tacto de los operarios de la Bodega. Sin dejar de olvidar, que estas características son las que le confieren las aptitudes físicas a la cartulina para satisfacer las necesidades de los clientes de la empresa. Estas características están preestablecidas por la industria, mientras que sus valores son parte del know-how de la empresa. La tabla 4, muestra algunas de las características que serán los candidatos para nuestros atributos difusos.

---

[4] La figura 6 muestra la interfaz de consulta de FSQL.
[5] Esas *etiquetas lingüísticas* las llamamos {rango mínimo, normal, rango máximo} en el modelo conceptual de nuestra BDRD.

| Características de una PILA | Características de un ROLLO |
|---|---|
| Pliegos parejos y planos. | Bobinado parejo. |
| Pliegos planos. | Limpio. |
| Tarima seca y con dimensiones correctas. | Empalmes bien pegados e identificados. |
| Cantidad exacta. | Corte no pelusiento. |
| Formato especificado. | Formato especificado. |
| Identificación correcta. | Identificación correcta. |

**Tabla 4** Características de *pilas* y *rollos*

La representación de sus atributos difusos es como sigue:

**Entidad Cartulinas Estucadas**: Definida por el esquema {*Cod_cart*, *Tono_Cara*, *Tono_Reverso*}. En que *Cod_cart* corresponde al código de la cartulina, por lo tanto es un dato no *fuzzy*. Mientras que *Tono_Cara* y *Tono_Reverso* son atributos de Tipo 3, por ser de dominios subyacentes no ordenados con las etiquetas lingüísticas {blanco, amarillo, café, manila}. En efecto es posible definir una relación de semejaza entre ellos. Estos atributos hacen referencia al color por ambos lados de la cartulina.

**Entidad Pilas**: Definida por el esquema {*Cod_Pila*, *Formato_Largo*, *Formato_Ancho*, *Estado*}. En que *Cod_Pila* corresponde al código de la pila, y por tanto es un dato no difuso. Mientras que *Formato_Largo* es un atributo de Tipo 2, definido por las etiquetas lingüísticas {corta, óptima, larga, muy larga} (ver figura 3), de igual forma *Formato_Ancho* es un atributo de Tipo 2, definido por las etiquetas lingüísticas {angosto, ancho, muy ancho}, ya que en todos esos casos se trata de atributos cuyo dominio es ordenado. El atributo *Estado* es considerado de Tipo 3, por ser de dominio subyacente no ordenado con las etiquetas lingüísticas {golpeado, mojado, orilla picada, englobado, sucio, picaduras, rayas en la superficie}. En efecto, es posible definir una relación de similitud entre ellas (similar a la tabla 5, pero para rollos).

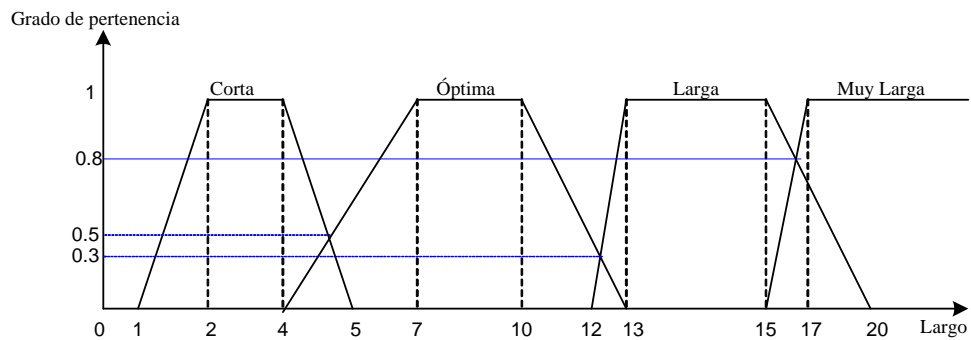

**Fig. 3** Función de pertenencia de las etiquetas lingüísticas del atributo "Formato_Largo"

**Entidad Rollos:** Definida por el esquema {*Cod_Rollo*, *Formato_Largo*, *Formato_Ancho Altura*, *Peso*, *Estado*}. En que *Cod_Rollo* corresponde al código del rollo, y por tanto es un dato no difuso. Mientras que *Formato_Largo* y *Formato_Ancho* corresponden al largo en metros de cartón enrollado, y al diámetro en centímetros del rollo, en ambos casos se trata de un atributo de Tipo 2, definido por las etiquetas lingüísticas {rango mínimo, normal, rango máximo}. *Altura* es un atributo de Tipo 2, definido por las etiquetas lingüísticas {baja, mediana, alta}. *Peso* es un atributo de Tipo 2, definido por las etiquetas lingüísticas {bajo, optimo, sobre}. *Estado* es un atributo considerado de Tipo 3, por ser de dominio subyacente no ordenado con las etiquetas lingüísticas {englobado, deslaminado, húmedo, sucio, rayas, curvas, empalme defectuoso, orilla crespa, disparejo}. En efecto, es posible definir una relación de similitud entre ellas, tal como lo muestra la tabla 5.

| $s_r$ | Englob. | Deslamin. | Hum. | Sucio | Rayas | Curvas | Emp. Defect. | Orilla Crespa | Disparejo |
|---|---|---|---|---|---|---|---|---|---|
| Englobado | 1 | 0 | 0 | 0 | 0 | 0,3 | 0,5 | 0,6 | 0 |
| Deslamin. | 0 | 1 | 0 | 0 | 0 | 0 | 0,8 | 0 | 0,1 |
| Humedad | 0 | 0 | 1 | 0 | 0 | 0 | 0 | 0 | 0 |
| Sucio | 0 | 0 | 0 | 1 | 0,8 | 0 | 0 | 0 | 0 |
| Rayas | 0 | 0 | 0 | 0,8 | 1 | 0 | 0 | 0 | 0 |
| Curvas | 0,3 | 0 | 0 | 0 | 0 | 1 | 0 | 0,8 | 0,5 |
| Emp. Defect. | 0,5 | 0,8 | 0 | 0 | 0 | 0 | 1 | 0,1 | 0,3 |
| Orilla Crespa | 0,6 | 0 | 0 | 0 | 0 | 0,8 | 0,1 | 1 | 0,8 |
| Disparejo | 0 | 0,1 | 0 | 0 | 0 | 0,5 | 0,3 | 0,8 | 1 |

**Tabla 5** Función de similitud para atributo "estado"

Esta descripción de los atributos difusos, permite hacer la siguiente consulta a la BDRD en FSQL, relativa a la Entidad Cartulinas Estucadas:

SELECT cartulina.% FROM cartulina WHERE tono_cara FEQ $blanco THOLD 0.5 and tono_reverso FEQ $blanco THOLD 0.5[6];

Que en lenguaje natural significa:

*"Obtenga todos los datos de las cartulinas estucadas que su tono de cara sea posiblemente igual a blanco (en grado mínimo 0.5) y que el tono reverso también sea posiblemente igual a blanco (en grado mínimo 0.5)".*

La figura 4, muestra la interfaz de FSQL y el resultado de la consulta[7].

**Fig. 4** Resultado de la consulta

---

[6] FEQ es un comparador de posibilidad, llamado "Fuzzy Equal"; $blanco es una etiqueta lingüística; y THOLD es usado para indicar un umbral de cumplimiento.

[7] CDEG(…) corresponde a un *grado de posibilidad* $\mu_A(u)$, donde *A* es un conjunto difuso y *u* es un elemento.

Mientras que la figura 5, muestra la Entidad Cartulinas Estucadas.

**Fig. 5** Entidad Cartulinas Estucadas

## 6   Conclusiones

Este trabajo nos ha permitido discutir, con múltiples ejemplos de la vida diaria, además de un caso real, la descripción de atributos difusos, bajo un modelo relacional difuso, llamado GEFRED (GEneralized model for Fuzzy RElational Databases). Esto ha permitido extender el esquema tradicional de los sistemas de información, a saber: <objeto, atributo, valor>, a la información imprecisa (en su representación) e incierta (en su consulta) con la teoría de conjuntos difusos. El nuevo esquema, lo podemos representar como: <objeto, atributo difuso, valor>. Donde, los atributos difusos del modelo relacional difuso GEFRED, se encuentran clasificados en: Tipo 1 (valores precisos), Tipo 2 (valores imprecisos sobre referencial ordenado), y Tipo 3 (valores imprecisos sobre referencial no ordenado). Mientras que los valores, pueden ser: valores numéricos (CRISP), escalares simples (LABEL), etiquetas lingüísticas (TRAPECIO), valores parcialmente desconocidos entre dos valores precisos (INTERVALO), valores parcialmente desconocidos entre un valor preciso (APROXIMADAMENTE), valores desconocidos (UNKNOWN), valores inaplicables (UNDEFINED), valores nulos (NULL), valores excluyentes de números o escalares (SIMPLE), y relaciones de similitud (DISTRIBUCION POSIBILIDAD).

En que cada uno de ellos, es visto como un conjunto difuso, para los cuales existe un grado de pertenencia o grado de posibilidad. El sistema FSQL, permite calcular este grado por medio de la función CDEG(…).

El caso de estudio ha permitido verificar que el tiempo de respuesta de la consulta es de 0.7 segundos, mientras que para transformar la sentencia FSQL, en cuestión, en sentencia SQL, demoró 1,8 segundos. Lo que es una enormidad, si se piensa que sólo se tienen 14 tuplas (ver figura 5). Esto es debido a que cada tabla que describe los atributos Tipo 2 y Tipo 3 (ver tablas

2 y 3), debe ser traducida a otras tablas, que contienen los valores de los atributos consultados. La FMB (Fuzzy Metaknowledge Base) contiene siete tablas de conversión de atributos difusos Tipo 2 y Tipo 3. Los detalles del diseño lógico de estas tablas se encuentran en [5].

Esto nos hace pensar que la simulación de la imprecisión y la incertidumbre en una base de datos relacional difusa, solo es posible hacerlo en un sistema de información cartesiano (0 o 1), lo que implica la creación de múltiples tablas de conversión para simular el hecho que un valor se encuentra en el intervalo [0,1].

Por lo tanto, la interfaz FSQL de la figura 4, mejora, sin lugar a duda, la calidad de la información, pero el costo en tiempo CPU, por una parte, de almacenamiento de datos difusos en una BDR, y por otra parte, de transformación de las consultas FSQL en SQL, sigue siendo elevado.

Esta constatación, nos ha permitido cuestionar las soluciones tecnológicas orientadas a la *gestión del conocimiento* [7], por medio de la *gestión de la información* y la *gestión de los datos*.

# Referencias